\documentclass[12pt]{iopart}
\usepackage{graphicx}
\begin{document}

\title{Nuclear $\beta^-$-decay half-lives for $fp$ and $fpg$ shell nuclei}
\author{ Vikas Kumar$^{1}$, P.C. Srivastava$^{1}$ and Hantao Li$^{2}$}
\address{$^{1}$Department of Physics, Indian Institute of Technology,
  Roorkee - 247667, India}
\address{$^{2}$School of Science, North University of China, Taiyuan 030051, People Republic of China}
\ead{vikasphysicsiitr@gmail.com; pcsrifph@iitr.ac.in}

%%%%%%%%%%%%%%%%%%%%%%%%%%%%%%%%%%%%%%%%%
\begin{abstract}
In the present work we calculate the allowed $\beta^-$-decay half-lives of nuclei with $Z = 20 -30$ and N $\leq$ 50 systematically under the framework of the nuclear shell model.
A recent study shows that some nuclei in this region belong to the island of inversion.
We perform calculation for $fp$ shell nuclei
using KB3G effective interaction. In the case of Ni, Cu, and Zn, we used JUN45 effective interaction.
Theoretical results of $Q$ values, half-lives,
excitation energies, log$ft$ values, and branching fractions are discussed and compared with the experimental data.
In the Ni region, we also compared our calculated results with recent experimental data [Z. Y. Xu {\it et al.},
\emph{Phys. Rev. Lett.} \textbf{113}, 032505, 2014]. Present results agree with the experimental
data of half-lives in comparison to QRPA.
\end{abstract}
\pacs{21.60.Cs -  shell model, 23.40.-s -$\beta$-decay}
%Keywords: $\beta$-decay, shell model\\
%\submitto{\JPG}

\maketitle

\vspace{1cm}
%%%%%%%%%%%%%%%%%%%%%%%%%%%%%%%%%%%%%%%%%%%%%%%%%%%%%%%%%%%
\section{Introduction}

The neutron density becomes very diffused and the single-particle spectrum shows the similarity of the harmonic-oscillator as we approach
towards the neutron-drip line \cite{dob,brown}. We can see this effect at $N=40$ shell gap. 
Recently intruder configurations are found in the neutron rich nuclei around this shell gap \cite{prc8161301,prc8564305,prc8614325,prl102142501,prl115192501}.
Thus the nuclear structure study of these nuclei in this region is very important \cite{p3104,prc8254301,prc9014302}.

Sorlin {\it et al.} \cite{sorlin1}, observed the experimental beta decay half-lives of neutron rich
$^{57-59}$Ti, $^{59-62}$V, $^{61-64}$Cr, $^{63-66}$Mn, $^{65-68}$Fe and $^{67-70}$Co
isotopes around $N=40$ at GANIL. Sorlin {\it et al.},  also compared experimentally observed half-lives of V and Cr isotopes with QRPA calculation.
The QRPA calculations for $^{58,59}$Cr, $^{61-63}$Mn, $^{63-64}$Fe, $^{65-67}$Co and $^{67,69}$Ni with folded-Yukawa single-particle potential are
reported in ref. \cite{moller}. The authors of \cite{moller}, have also compared the results with Bender {\it et al.,} \cite{bender} based on Nilsson potential.
Using RIBF facility at RIKEN, the half-lives of twenty neutron-rich nuclei with $Z=27-30$ are reported in ref. \cite{Xu}. In that work,
sizable magicity was reported for both the proton number $Z=28$ and the neutron number $N=50$ in $^{78}$Ni. 
A sudden shortening of half-lives of the nickel isotopes beyond $N=50$ were observed in this work, although this effect is not found in the Cu-Ge-Ga chains.
The half-lives results from LISE2000 spectrometer at GANIL for $^{71}$Co and $^{73}$Co are reported in ref. \cite{Sawicka}.
The half-lives of $^{77,78}$Ni were measured for the first time by Hosmer {\it et al.} \cite{hosmer}, at NSCL, MSU.
Recently the beta decay half-lives of 38 very neutron-rich Kr to Tc isotopes are measured at RIKEN \cite{Nishimura}.

Despite the progress in the experimental side, we need theoretical estimates for half-lives of neutron rich nuclei, especially those belonging to the island of inversion. 
These calculations are based on allowed GT-transitions. Many theoretical calculations from the quasiparticle random phase
approximation (QRPA) based on the Hartree-Fock Bogoliubov theory \cite{q1} or other global models \cite{q2,q3,q4} are available in the literature.
These calculations underestimate the correlation among nucleons which predict GT-strength at low-energies.
%%The shell model results are able to predict half-lives close to the experimental data for the neutron rich 
%%nuclei \cite{Zhi,s1,s2,s3}.
Recently, shell model calculations for the $\beta^-$-decays of $Z=9-13$ nuclei are reported by Li and Ren \cite{li}.
The motivation of the present work is to study $\beta^-$-decay properties of $Z = 20 - 30$ nuclei using the nuclear shell model.

This paper is organized as follows. In section 2, we present the formulas for the calculation of
$\beta^-$ decay half-lives. Shell model spaces, effective interactions and the quenching factors adopted
in our calculations are reported in section 3. In section 4, we present theoretical results along with the experimental data wherever 
available. Finally, summary and conclusions are drawn in section 5.

%%%%%%%%%%%%%%%%%%%%%%%%%%%%%%%%%%%%%%%%%%%%%%%%%%%%%%%%%%%

\section{Formalism}

In the beta decay, the transitions start from the ground state of the parent nuclei to different excited states (only those inside the energy window defined by $Q$-value)
of the daughter nuclei according to the selection rule of beta -- decay.
The $ft$ value is calculated by
\begin{equation}
 ft = \frac{6177}{[(g_A)^2 B(GT) + B(F)]}
\end{equation}
where, $g_A$ (= 1.260) is the axial-vector coupling constant of the weak interactions. Here, $f$ is a phase-space integral that contains the lepton kinematics. The $B(GT)$ and $B(F)$ are the Gamow-Teller and Fermi
matrix elements. The total half-life is calculated as
\begin{equation}
 t_{1/2}= ({\sum_i {\frac{1}{t_i}}})^{-1}
\end{equation}
 where $t_{i}$ is the partial decay half life of the daughter's state i.
 %The nucleus $\beta$ -decay half-life $t_{1/2}$ is determined by all these partial half-lives $(t_{i})$.\\
The partial half-life of the allowed $\beta^-$-decay is given by \cite{suhonen}, 
\begin{equation}
 t_{i}= 10^{logft - logf_A}
\end{equation}
here, ${f_A}$ is the Gamow -Teller (axial-vector) phase space factor. Because the $ft$ values are usually large, they are normally 
expressed 
in terms of `log $ft$ values'. The log $ft$ value is defined as log$ft$ $\equiv$ log$_{10}(f_At_{i}[s])$.

The partial half-life $t_i$ is related to the total half-life $t_{1/2}$ of the allowed $\beta^-$-decay as
\begin{equation}
 t_{i}= {\frac{t_{1/2}}{b_r}}
\end{equation}
where,  ${b_r}$ is the branching ratio for the level with partial half-life ${t_i}$. The
$B(GT)$ is the Gamow-Teller matrix element
\begin{equation}
 B(GT)= (\frac{g_A}{g_V})^2{\langle{\sigma\tau}\rangle}^2
\end{equation}
%where, $\frac{g_A}{g_V}$ = -1.264 $\pm$ 0.002 \cite{Dubbers}.
The nuclear matrix element of Eqn. (5) for the Gamow-Teller operator is
\begin{equation}
 {\langle{\sigma\tau}\rangle} = {\langle {f}|| \sum_{k}{\sigma^k\tau_{\pm}^k} ||i \rangle}/\sqrt{2J_i + 1},
\end{equation}
where $f$ and $i$ refer to all the quantum numbers needed to specify the final and initial states, respectively, $\pm$ refers to $\beta^{\pm}$ decay,
$\tau_{\pm} = \frac{1}{2}(\tau_x + i\tau_y)$ with $\tau_+p$ = $n$, $\tau_-n$ = $p$, and $J_i$ is the total angular momentum of the initial-state. The sum in Eqn. (6) runs over all the nucleons.
We calculate the values of ${f_{A}}$ and ${\langle{\sigma\tau}\rangle}^2$ from the refs. \cite{wilkinson,wilkinson B.E,wilkinson A.G}.

We also report the $\beta$-decay $Q$ value using shell model calculations, the theoretical $\beta$-decay $Q$ value is given by
\begin{equation}
Q = (E(SM)_i + E(C)_i) - (E(SM)_f + E(C)_f)
\end{equation}
where $E(SM)$ is the nuclear binding energy of the interaction of the valence particles among themselves, which can be evaluated from the shell model calculation,
$E(C)$ is the valence space Coulomb energy, and subscripts $i$ and $f$ denote the parent and daughter nuclei, respectively.
The expression for $E(C)$ is taken from ref. \cite{Caurier}.

\section{\bf{Hamiltonian and quenching factor}}

In the present work we performed calculation for $fp$ and $f_{5/2}pg_{9/2}$ shell nuclei using the shell model code NuShellX@MSU \cite{MSU-NSCL}.
For $fp$ shell nuclei,
we used KB3G effective interaction \cite{kb3g}. In the case of $f_{5/2}pg_{9/2}$ model space for  Ni, Cu and Zn isotopes, we performed calculations with
JUN45 effective interaction \cite{jun45}. 
 While NuShellX is a set of shell model computer codes written by Bill Rae \cite{rae}, NuShellX@MSU is a set of wrapper codes written 
by Alex Brown. NushellX is based on OpenMP to use many cores with high efficiency for Lanczos interactions. NuShellX uses a proton-neutron basis. With NuShellX, it is possible 
to diagonalize J-scheme matix dimensions up to $\sim$ 100 million. The Hamiltonian is written as a sum of three terms $H =H_{nn}+H_{pp}+H_{pn}$. With
this code it is possible to calculate spectroscopic factors, two-nucleon transfer amplitudes and one-body transition densities. 
 In the DENS program, it is possible to use radial wavefunctions from harmonic-oscillator, 
Woods-Saxon or Skyrme-density functions methods for the matrix elements. With the NushellX code it is possible to calculate log$ft$ values and $GT$ strength 
to individual final states but not the total strength. The parameters which we need for the calculations are the model space and the corresponding effective interaction (two-body
matrix elements).

 We performed calculation for nuclei in two different model spaces.  Thus we used two different effective interactions JUN45 and KB3G.
The KB3G \cite{kb3g} interaction is extracted from the KB3 interaction by introducing mass dependence and
refining its original monopole changes in order to treat properly the $N=Z=28$ shell closure
and its surroundings.  In order to recover simultaneously the good gaps around $^{48}$Ca and $^{56}$Ni, here $T=0$ and $T=1$ modifications are different.
The single-particle energies for KB3G effective interaction are taken to be -8.6000, -6.6000,
-4.6000 and -2.1000 MeV for the $f_{7/2}$, $p_{3/2}$, $p_{1/2}$ and $f_{5/2}$ orbits, respectively. 

The JUN45 effective interaction, which was recently developed by Honma \cite{jun45}, is a realistic interaction based on the Bonn-C potential
fitting by 400 experimental binding and excitation energy data with mass numbers A = 63 - 96. 
 For the JUN45 interaction, the single-particle energies are taken to be -9.8280, -8.7087,
-7.8388 and -6.2617 MeV for the $p_{3/2}$, $f_{5/2}$ , $p_{1/2}$ and $g_{9/2}$ orbits, respectively. 
The large number of experimental data around $N=50$ shell closure has been taken for the fitting of this interaction.
For this interaction an rms deviation is 185 keV.
 We performed full-fledged calculation, but for few $fp$ shell nuclei
we freeze 2 - 8 neutrons in the $f_{7/2}$ orbital.

Following Ref. \cite{brownrgt} we can define matrix elements, $M(GT)$,  in terms of reduced transition probability, $B(GT)$, by following
\begin{equation}
M(GT)= [(2J_{i}+1)B(GT)]^{1/2},
\end{equation}
this is independent of the direction of the transitions. Here $J_{i}$ is the total angular momentum  of the initial state.
To get effective $g_{A}$, first we normalize the $M(GT)$
to the ``expected" total strength W, defined by

\begin{equation}
  W=\left\{
  \begin{array}{@{}ll@{}}
    |g_{A}/g_{V}|[(2J_{i}+1)3|N_{i}-Z_{i}|]^{1/2} , & for N_{i} \neq  Z_{i},\\
    |g_{A}/g_{V}|[(2J_{f}+1)3|N_{f}-Z_{f}|]^{1/2} , & for N_{i} = Z_{i},
  \end{array}\right.
\end{equation}

The matrix elements $R(GT)$ are defined as
\begin{equation}
R(GT) = M(GT)/W.
\end{equation}

The comparison of the experimental versus the theoretical $R(GT)$ values are plotted in Fig. 1.
The $R(GT)_{exp}$ values are taken from the experimental log$ft$ values (as given in the table 1).

\begin{figure}
\begin{center}
\includegraphics[width=6.2cm]{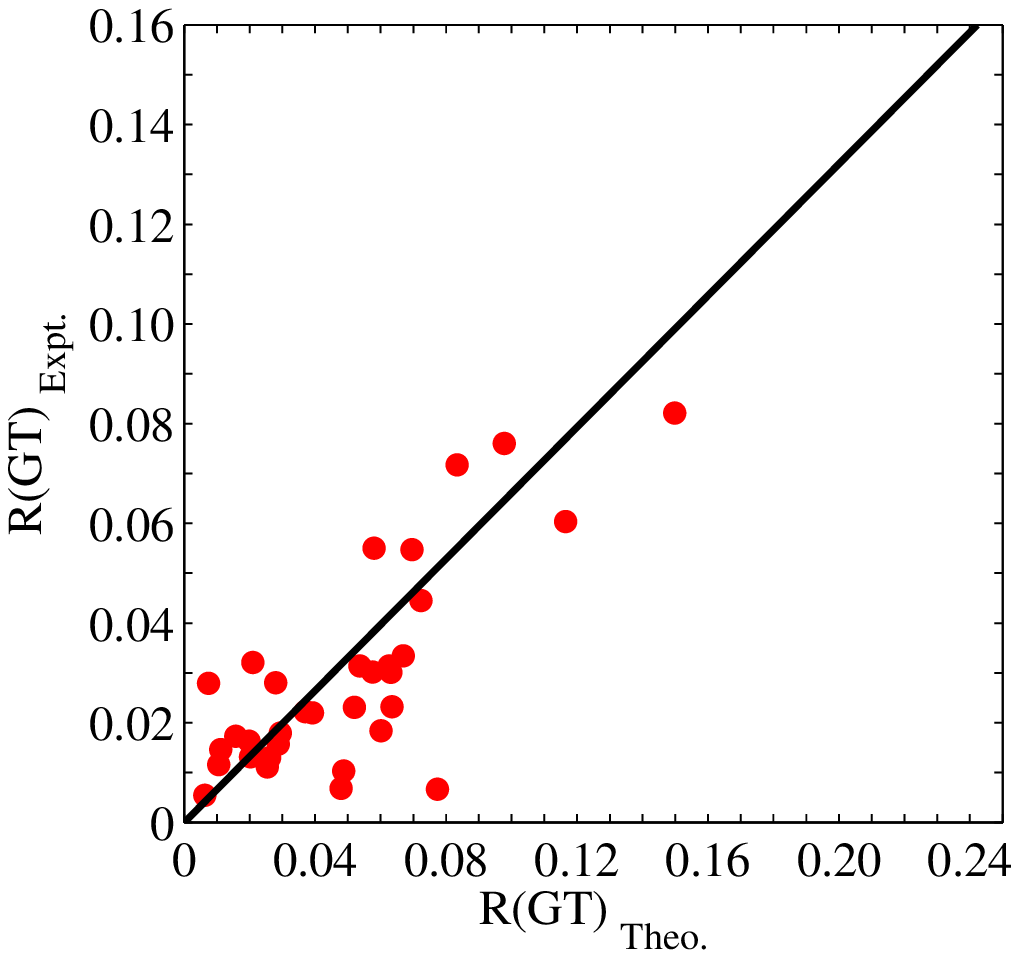}
\includegraphics[width=6.7cm]{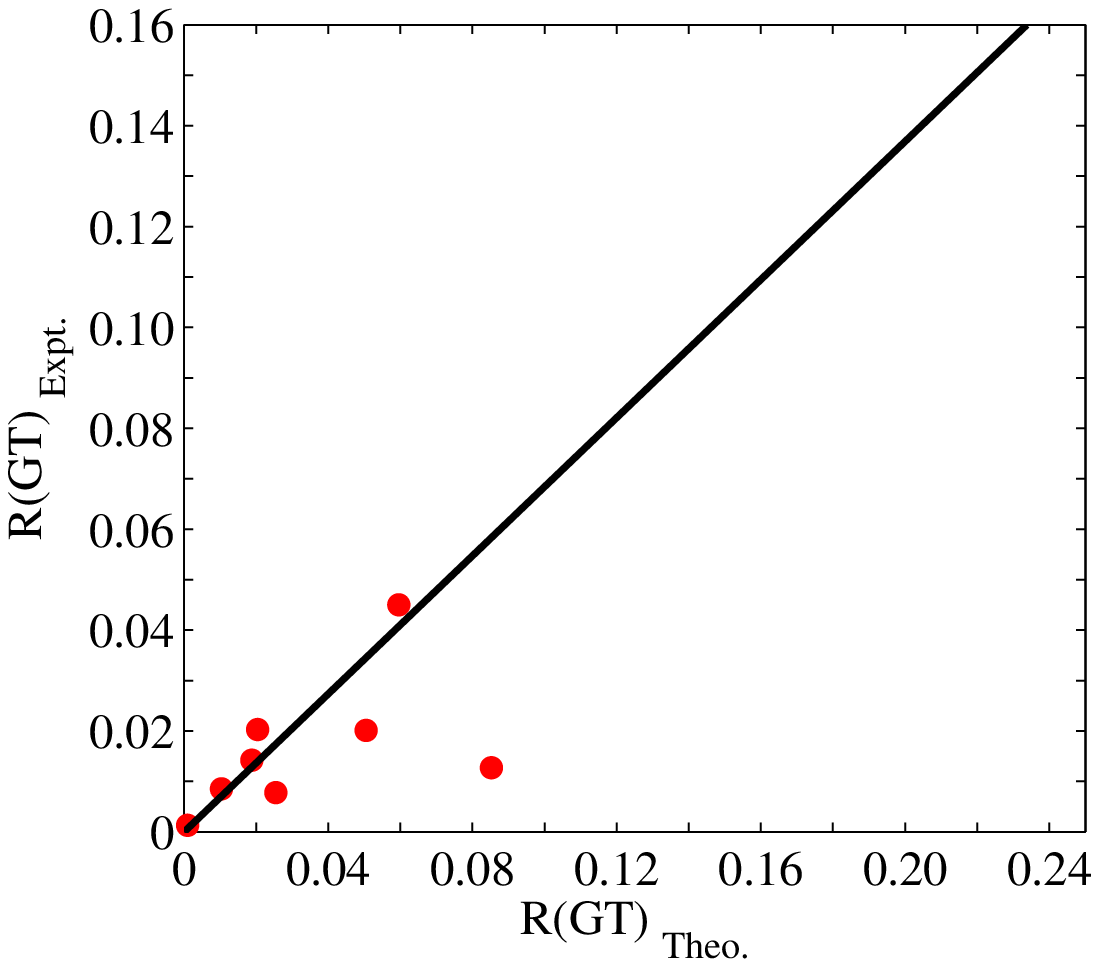}
\caption{ Comparison of the experimental matrix element $R(GT)$ values with theoretical calculations based
on the ``free-nucleon" Gamow-Teller operator. Each transition is indicated by a point in the x-y plane.
Theoretical and experimental value given by the x, y coordinates, respectively. Left panel for $fp$ model space with KB3G effective
interaction, while the right panel for $f_{5/2}pg_{9/2}$ model space with JUN45 effective interaction.}
\end{center}
\end{figure}
 The quenching factor is defined as the square root of the ratio of the experimental measured rate to the calculated
rate in a full $0\hbar\omega$ calculation. Because the observed Gamow-Teller strength appears to be systematically smaller than what
is theoretically expected on the basis of model independent Ikeda sum rule ``3(N-Z)". 
The quenching factor $q$ in a given model space is obtained by averaging all the ratios between the experimental
and theoretical $R(GT)$ values.
The points follow nicely a straight line whose slope gives the average quenching factor.
In the present work we performed the shell model 
calculations for two different model spaces. Thus, we have two different quenching factors.
From this work for $pf$ shell nuclei we have $q$ = 0.660 $\pm$ 0.016, while for
$f_{5/2}pg_{9/2}$ shell nuclei
we have $q$ = 0.684 $\pm$ 0.015.

\begin{table}
\begin{center}
\caption{ Comparisons of the theoretical log$ft$ values, excitation energies, and branching
percentages of $\beta$-decays of the concerned nuclei with the experimental values. An asterisk beside
the experimental excitation energy indicates that the experimental $J^\pi$ of this state is uncertain.
References to the experimental data are given in the last column.}
\begin{tabular}{r@{\hspace{4pt}}c@{\hspace{2pt}}c@{\hspace{2pt}}c@{\hspace{2pt}}c@{\hspace{2pt}} c@{\hspace{2pt}}c@{\hspace{2pt}}c@{\hspace{2pt}}c@{\hspace{2pt}}c@{\hspace{2pt}}c@{\hspace{2pt}} c@{\hspace{2pt}}c@{\hspace{2pt}}c@{}}
\hline
&    &&\multicolumn {2} {c} {Ex. energy (keV)} &&\multicolumn {2} {c} {log$ft$ value}&& \multicolumn {2} {c} {Branching ($\%$)}\\
\cline{4-5}
\cline{7-8}
\cline{10-11}
$^{A}$Z$_{i}(J^{\pi})$  & $^{A}$Z$_{f}(J^{\pi})$  &&
      \multicolumn{1}{c}{Theo.}&\multicolumn{1}{c}{Expt.} &&
      \multicolumn{1}{c}{Theo.}& \multicolumn{1}{c}{Expt.} &&
      \multicolumn{1}{c}{Theo.}&\multicolumn{1}{c}{Expt.}&&
      Ref.\\

\hline
$^{52}$Ca($0^+$)   &  $^{52}$Sc($1^+$) && 1306& 1636 && 4.309 &5.07  &&93.81   &   86.8&&  \cite{yang}\\
                   &                   && 2606&4266  && 4.997 & 5.8  &&6.19   &1.4    &&         \\
$^{54}$Ca($0^+$)   & $^{52}$Sc($1^+$)  &&0    & 247  &&3.921  &  4.25&& 89.25 &97.17  && \cite{mantica} \\
$^{54}$Sc($3^+$)   & $^{54}$Ti($4^+$)  && 2452&2497  && 5.173 & 5.3  && 50.26 & 33    && \cite{crawford} \\
                   & $^{54}$Ti($2^+$)  && 1285& 1495 && 5.731 & 5.7  && 22.53 & 21    &&         \\
$^{55}$Sc($7/2^-$) & $^{55}$Ti($9/2^-$)& &1562&2146  && 4.660 &5.3   && 10.84  & 11    && \cite{crawford} \\
                   & $^{55}$Ti($7/2^-$)&& 1079&1796  && 4.341 & 5.0  &&28.09  & 22    &&         \\
                   & $^{55}$Ti($5/2^-$)&&0    &592   && 4.202 &5.0   && 61.1 &39     &&         \\
$^{57}$Ti($5/2^-$) & $^{57}$V($7/2^-$) && 161  & 175*   && 4.259 & 5.7$\pm$0.2 && 51.28  & 5$\pm$3    && \cite{liddick} \\
                   &                   && 1836 & 1754*  && 6.806 &4.9$\pm$0.2  &&0.06    & 16$\pm$2   &&         \\
                   &                   && 2328 & 2476*  && 4.654 &5.0$\pm$0.2  && 6.75   & 7.3$\pm$0.7&&         \\
                   & $^{57}$V($3/2^-$) && 1711 & 1732*  && 5.141 & 6.0$\pm$0.2 && 3.11  & 1.1$\pm$0.7&&         \\
                   &                   && 2040 & 2036*  && 5.499 &4.8$\pm$0.2  && 1.14  &16$\pm$2    &&         \\
$^{56}$V($1^+$)    & $^{56}$Cr($2^+$)  && 920  & 1006   && 5.477 & 5.63      && 2.67   & $<$4       && \cite{pfmantika}\\
                   &                   && 1705 & 1830   && 5.103 &6.01         && 3.95   &1.0         &&         \\
                   &                   && 2178 & 2324   && 5.384 &5.87         && 1.53   & 1.0        &&         \\
                   & $^{56}$Cr($0^+$)  && 0    &0       && 4.399 &4.62         &&  52.22 & 70         &&         \\
                   &                   && 1463 & 1675   &&4.242  &4.63         && 33.30  &   26       &&         \\
$^{57}$V($3/2^-$)  & $^{57}$Cr($5/2^-$) && 195   & 268*  && 5.404  & 4.67         && 7.25  & 47         &&\cite{pfmantika}\\
                   &                    && 692   & 693*  && 8.785  & 4.93         && 0.002 & 20         &&      \\
                   &                    && 1616  & 1583* && 6.814  & 5.51         && 0.11  &  3         &&      \\
$^{58}$V($1^+$)    & $^{58}$Cr($2^+$)   && 731   & 879   && 4.047  & $>$5.3         &&44.2  & $<$34      &&\cite{caroline_DN}\\
                   & $^{58}$Cr($0^+$)   && 0     & 0     && 4.089  &  $>$5.3        && 54.65 & $<$38        &&         \\
$^{60}$Cr($0^+$)   & $^{60}$Mn($1^+$)   &&   0   & 0     &&3.485   & 4.2$\pm$0.1  && 95.26 &88.6$\pm$0.6&& \cite{S. N. Liddick} \\
                   &                    &&1120   & 759   && 4.469  & 5.0$\pm$0.2  && 4.14  & 5.0$\pm$0.5&&         \\
$^{61}$Cr($5/2^-$) & $^{61}$Mn($7/2^-$) && 185   & 157   && 4.887  & 5.6$\pm$0.1  &&  13.62&  9$\pm$2   && \cite{H. L. Crawford} \\
                   & $^{61}$Mn($5/2^-$) && 1645  &1497*  && 6.484  & 4.9$\pm$0.2  && 0.15  & 20$\pm$5   &&         \\
                   &                    && 2098  &2032*  && 4.356  & 5.4$\pm$0.2  && 15.46 & 5$\pm$1    &&         \\
                   & $^{61}$Mn($3/2^-$) && 866   & 1142* && 5.559  & 5.6$\pm$0.4  && 2.01  & 5$\pm$4    &&         \\
                   &                    && 1623  & 1860* && 7.026  & 4.8$\pm$0.1  && 0.045 & 20$\pm$2   &&         \\
$^{62}$Cr($0^+$)   & $^{62}$Mn($1^+$)  && 579    & 0     && 3.782  & $\sim$4.2     && 43.78  & $\sim$72    && \cite{Alan L. Nichols}\\
                   &                   && 823    & 640   && 3.631  & $\sim$4.4     && 56.14  & $\sim$25    &&  \\
                   &                   &&1794    & 1500* && 6.291  & 5.3           && 0.08  & 3           &&  \\
\hline
\end{tabular}
%\label{be2}
\end{center}
\end{table}

\addtocounter{table}{-1}

\begin{table}
  \begin{center}
    \leavevmode
    \caption{{\em Continuation.\/}}
\begin{tabular}{r@{\hspace{4pt}}c@{\hspace{2pt}}c@{\hspace{2pt}}c@{\hspace{2pt}}c@{\hspace{2pt}} c@{\hspace{2pt}}c@{\hspace{2pt}}c@{\hspace{2pt}}c@{\hspace{2pt}}c@{\hspace{2pt}}c@{\hspace{2pt}} c@{\hspace{2pt}}c@{\hspace{2pt}}c@{}}
\hline
&    &     & \multicolumn {2} {c} {Ex. energy (keV)} &&\multicolumn {2} {c} {log$ft$ value}&& \multicolumn {2} {c} {Branching ($\%$)} & \\
\cline{4-5}
\cline{7-8}
\cline{10-11}
$^{A}$Z$_{i}(J^{\pi})$  & $^{A}$Z$_{f}(J^{\pi})$  &&
      \multicolumn{1}{c}{Theo.}&\multicolumn{1}{c}{Expt.} &&
      \multicolumn{1}{c}{Theo.}& \multicolumn{1}{c}{Expt.} &&
      \multicolumn{1}{c}{Theo.}&\multicolumn{1}{c}{Expt.}&& Ref.\\

\hline
$^{60}$Mn($1^+$)   & $^{60}$Fe($2^+$)  && 658    & 823   && 4.866  & 5.6$\pm$0.2   &&  7.35  & 4.2$\pm$1.2 && \cite{m.p carpenter}\\
                   & $^{60}$Fe($0^+$)  && 0      & 0     && 4.070  & 4.46$\pm$0.04 &&  66.98 & 88$\pm$2    && \cite{m.p carpenter}\\
                   &                   && 1345   & 1974  && 4.242  & 5.15$\pm$0.07 &&   20.15&  5.0$\pm$0.6&&  \\
                   &                   && 2291   & 2356  && 4.860  & 5.3$\pm$0.1   &&   2.51  & 3.0$\pm$0.5 &&  \\
$^{61}$Mn($5/2^-$) &$^{61}$Fe($7/2^-$) && 991   & 960   && 5.995  & 6.6(1)        &&0.35    &0.49(7)      && \cite{D. Radulov}\\
                   &                   && 1906   & 1929* && 7.532 & 5.40(2)       &&  0.05  &   2.7(1)    &&  \\
                   &                   && 2120   & 2144* && 5.990  & 5.43(2)       && 0.14  &    1.8(1)   &&  \\
                   &$^{61}$Fe($5/2^-$) && 243    & 207   && 4.575 & 5.38(13)      && 15.95  & 12.6(8)     && \cite{D. Radulov}\\
                   &                   && 1149   &1161   && 5.681 & 5.64(1)       && 0.65   & 3.25(7)     &&   \\
                   &$^{61}$Fe($3/2^-)$  && 0      & 0     && 4.052  & 5.02(3)       &&62.25   & 33(1)       && \cite{D. Radulov}\\
                   &                   && 831    & 629   && 4.291 & 4.77(1)       &&  20.36 & 39.0(6)     &&  \\
                   &                   && 1464   & 1253  && 6.892 & 6.4(5)        &&  0.03  & 0.57(6)     &&  \\
$^{62}$Mn($4^+$)   & $^{62}$Fe($5^+$)  && 3719   & 3714* && 5.629  & $>$6.4        &&   1.11 & 1.2         && \cite{NDS113}\\
                   & $^{62}$Fe($4^+$)  && 2111   & 2177  && 5.788  & 5.9           &&   2.17  & 8.4         && \cite{NDS113}\\
                   & $^{62}$Fe($3^+$)  && 2357   & 2692  && 4.274  & 5.5           &&  61.11 & 17          &&  \cite{NDS113}\\
                   &                   && 3384   & 3360* && 7.436  & $>$6.5        &&  0.02  & 1.2         &&  \\
$^{64}$Mn($1^+$)   &$^{64}$Fe($2^+$)   && 1424   & 1444* && 4.705  & $\sim$5.6        &&   15.52 & $\sim$7.3       &&  \cite{NDS108} \\
                   &$^{64}$Fe($1^+$)   && 3384   & 3307* && 4.070  & $\sim$5.5          &&  25.66  & $\sim$4.0         &&  \\
                   &                   && 3910   & 4227* && 5.742  & $\sim$5.7       &&  0.40  & $\sim$1.4      &&  \\
$^{73}$Ni($9/2^+$) & $^{73}$Cu($9/2^+$) && 1812 & 1489&& 5.062 & $>$5.8  &&25.61 &$<$15.7&&\cite{S. Franchoo}\\
                   & $^{73}$Cu($7/2^+$) && 1423 & 961 && 4.866 & $>$5.9  && 40.02&$<$19.2&& \\
$^{76}$Cu($3^+$)   &$^{76}$Zn($4^+$)   && 1641 & 1296 && 5.769 & 5.8   && 11.78  &  24(4)    & & \cite {nds74}\\
                   &                   && 1814 & 2633*&& 5.467 & 5.9   && 21.99  &   10(2)    && \\
                   &                   && 2666 & 2814*&& 6.669 & 6.3   &&  0.95 &    3.8(7)  && \\
                   & $^{76}$Zn($3^+$)  && 1788 & 3232*&& 5.191 & 6.0   && 21.51  &  6.2(12)   && \\
                   &                   && 2107 & 3273*&& 5.689 &6.5(3) &&  11.64 &  1.7(10)   && \\
                   & $^{76}$Zn($2^+$)  && 2244 & 2266*&& 6.370 & 6.3(1)&& 2.28   &  4.9(13)   &&\\
                   &                   && 2540 & 2349*&&5.870  & 8(4)  && 6.34  & 0.1(9)     &&\\
$^{77}$Cu($5/2^-$) & $^{77}$Zn($5/2^-$)&& 1104 &1284* && 5.889 &6.5    && 4.53  & 2          &&\cite{N. Patronics}\\
                   &                   && 1813 &2082* &&6.003 & 6.1   &&2.64   & 3          && \\
$^{78}$Cu($5^-$)   & $^{78}$Zn($6^-$)  && 3732  & 3105 && 6.035 & 5.5(1)&& 13.09  &  21(3)     &&\cite{J. Van Roosbroeck}\\
$^{79}$Zn($9/2^+$) &$^{79}$Ga($11/2^+$)&& 2644  & 2561* && 5.054 & $\geq$6.3 &&  28.07  &   $\leq$2    &&\cite{NDS96}\\
                   &                   && 2656  & 2649* && 5.830 &  5.9 &&  4.67  &   4.0  && \\
                   &                   && 2846  & 2919*     && 3.782 &   5.7   &&  12.93  &   5.6  && \\
                   & $^{79}$Ga($9/2^+$)&& 2785  &  2741*&& 5.755 & 5.4      && 5.93  &   13    &&\\
                   & $^{79}$Ga($7/2^+$)&& 3105  & 3020* && 6.788 &  5.5 &&  0.54 &   7.4     &&\\
$^{80}$Zn($0^+$)   & $^{80}$Ga($1^+$)  &&617  & 686   && 3.746  & 5.6 &&  75.99 &   5.5  && \cite{R L Gill}\\
                   &                   && 844  & 965   && 4.202  & 5.2  && 23.12  &  10.1    &&\\
                   &                   &&1487   & 1504  && 5.444 & 5.6  &&  0.87  &   3.1    &&\\
                   &                   && 1674 & 2070  && 7.551 & 4.5  &&   0.006  &   20.5    &&\\
 \hline
\end{tabular}
%\label{be2}
\end{center}
\end{table}

\section{Results and discussions}

The comparison between theoretical and experimental log$ft$ values, excitation energies, and the branching percentages of  $\beta$-decays of the concerned
nuclei are shown in the table 1.
The theoretical and experimental excitation energies of each
state involved in the  $\beta$ - decays are listed in column 3 and 4, respectively. The theoretical and experimental
log$ft$ values are listed in  column 5 and 6, respectively.
The branching fractions are given by $b_r = t_{1/2}/t_{i}$, where $t_{1/2}$ is the nuclear $\beta$ - decays half life
and $t_{i}$ is the partial half-life of each transition. The theoretical branching fractions are listed in
column 7 and the experimental values are listed in column 8.  In this table, we present experimental
uncertain state  by `*'. The shell model results for $fp$ and $f_{5/2}pg_{9/2}$ shell nuclei are
showing a good agreement with the experimental data for excitation
energies, log$ft$ values and the branching ratios.

In the table 2, we compare the theoretical and the experimental  $\beta$-decay half-lives of the concerned nuclei.  The experimental $Q$ values, 
$\beta^-$-decay probabilities and the quenched theoretical sum $B(GT)$ values 
are also reported in this table.
The first and second columns are the parent and daughter nuclei, respectively.
 Column 3 presents the experimental $Q$ values, which are taken from \cite{wong}. In column 4 we present the sums of quenched $B(GT)$ values.
 Theoretical and experimental $\beta$-decays half-lives are presented in columns 5 and 6, respectively.
 In the last column we present the experimental probabilities of $\beta^-$-decay. The  probabilities  of $\beta^-$-decay
for most of the nuclei are 100\%. In this table, we present the experimental value
 which is unknown by '?'. 
 We determined the ground states using the shell model for the nuclei where the experimental ground states are not confirmed.
 In these cases we calculate the $\beta$-decay properties for the three lowest states as reported in table 1.

The $^{63-65}$Mn nuclei belong to the island of inversion \cite{prc8614325}. The experimental $\beta^-$-decay half-lives
for these three nuclei are 275 $\pm$ 4 ms, 90 $\pm$ 4 ms, and 84 $\pm$ 8 ms, respectively, while the calculated shell model
results are 216 ms, 61 ms and 73 ms, respectively. The calculations are remarkably close to the measured values.
 We also reported the calculated $\beta^-$-decay half-lives for some nuclei such as $^{57-58}$Ca, 
$^{59-61}$Sc, and $^{62}$Ti, where exact experimental values are still not known.

In the Fig. 2, we show the theoretical and experimental $\beta$-decay half-lives of concerned
nuclei taken from table 2.  We used the log frame to show $\beta$-decay half-lives. Here, the
$\beta$-decay half-lives decrease rapidly with the increasing neutron number.
In this figure we presented the experimental data with error bars while the theoretical results are
connected by solid lines with `+' sign. In the case of most $fp$ shell nuclei results are in a good agreement with experimental data.
For a few heavier even-even $fp$ shell nuclei (around $\sim N=40$), the calculated results are not in a good agreement with the experimental data.
This may be due to the missing $\nu d_{5/2}$ orbital in the model space. The results with JUN45 interaction is showing a good
agreement for Ni, Cu and Zn isotopes, except for 3 even-even Ni nuclei. We take new experimental results of half-lives from ref. \cite{Xu} for comparison with
the calculated values. Our results are closer to the experimental data in comparison to previously available QRPA results
in ref. \cite{q2}, because GT-strength is underestimated in QRPA.
Some of the concerned nuclei belong to the island of inversion, such as $\mathrm{^{63-65}Mn}$ and $\mathrm{^{67}Co}$, and their $\beta$-decays properties are well reproduced by our calculations.

The theoretical result shows better agreement with the experimental data for a relatively small neutron number.
As we move towards drip line, some of the theoretical results deviate from the experimental data.
This is because our calculations are unable to predict the ground state correctly, and uncertainties of the
$Q$ values are large. These facts are responsible for the large errors of the theoretical half-lives.

From the table 2, we also show the ratios between theoretical $\beta$-decay half-lives and
experimental data for the concerned nuclei in Fig. 3. Here, in the left figure we shown the results for $fp$ shell nuclei using KB3G
effective interaction, while in the right figure we shown the results of Ni, Cu and Zn isotopes using JUN45 effective interaction.

We also report the $\beta$-decay $Q$ values using Eqn. (7)
and compare with the experimental data in table 3. Although we are not using these theoretical $Q$ values
for the evaluation of $\beta$-decays half-lives as reported in table 1. We take the experimental
data from ref. \cite{wong} ( where `\#' indicates that the presented value is estimated from systematic trends).
The rms deviation between theory and experiment is 1162 keV.

In the Fig. 4, we show the comparison between the calculated and the experimental energy levels for $^{64}$Ni. In the case of
$^{64}$Co and $^{64}$Ni, we performed calculation in $fp$ model space using KB3G effective interaction. Here we put minimum 6
and 4 particles in $\pi f_{7/2}$ and $\nu f_{7/2}$ orbitals, respectively. The energy levels are in a good
agreement with the experimental data. The calculated partial half-life for $2_1^+$ to $0_1^+$ transition is 1.0 $ps$ while
corresponding experimental value is 1.088 $ps$. The calculated partial half-life for $2_2^+$ to $0_1^+$ transition is 1.11 $ps$,
although for this transition experimental data is not available.

\begin{table}
  \begin{center}
    \leavevmode
    \caption{Comparison of the theoretical $\beta$-decay half-lives
    with the experimental data for the concerned nuclei together with the experimental
$Q$ values \cite{wong}, $\beta^-$-decay probabilities and quenched theoretical sum $B(GT)$ values.}

\begin{tabular}{lrccccccccc}

% \begin{center}
\hline
   &    & $Q$ value & Sum &\multicolumn{2}{c}{Half-life}&& $\beta^-$\\
\cline{5-6}
%\cline{8-9}
$^{A}$Z$_{i}(J^{\pi})$  & $^{A}$Z$_{f}$  & (keV) &$B(GT)$&\multicolumn{1}{c}{Theo.}&\multicolumn{1}{c}{Expt.}&
   &$\%$\\
\cline{3-4}
%\cline{5-6}
%\cline{7-8}
\hline
$^{52}$Ca($0^+$)   &  $^{52}$Sc & 5900 &  0.1004  & 2.519 s &  4.6$\pm$0.3 s\cite{half} &          &100\\
$^{53}$Ca($1/2^-$) &  $^{53}$Sc & 9650\# & 0.140   & 268.87 ms &  461$\pm$90 ms\cite{half} &       &100 \\
$^{54}$Ca($0^+$)   &  $^{54}$Sc & 8820\# & 0.895   & 186.7 ms&  86$\pm$7 ms\cite{half}   &             &100 \\
$^{55}$Ca($5/2^-$) & $^{55}$Sc  & 16630\#&  0.0902  & 18.51 ms &    22$\pm$2 ms\cite{half} &          &100\\
$^{56}$Ca($0^+$)   & $^{56}$Sc  &10830\#&  0.799   & 1.56 ms  &     11$\pm$2 ms\cite{half}&            &100\\
$^{57}$Ca($5/2^-$) & $^{57}$Sc  & 13830\#&  0.322  & 3.90 ms & 5 ms($>$620 ns)\cite{half}&            &?\\
$^{58}$Ca($0^+$)   & $^{58}$Sc  & 12960\#&  1.386  &0.486 ms & 3 ms($>$620 ns)\cite{half}&               &?\\
$^{54}$Sc($3^+$)   & $^{54}$Ti  & 12000&  0.023  & 565.7 ms& 526$\pm$15 ms\cite{half} &         &100\\
$^{55}$Sc($7/2^-$) & $^{55}$Ti  & 11690&  0.221  & 51.30 ms & 96$\pm$2 ms\cite{half}   &         &100\\
$^{56}$Sc($1^+$)   & $^{56}$Ti  &14470\# &  0.489  & 22.37 ms & 26$\pm$6 ms\cite{half}   &            &100\\
$^{57}$Sc($7/2^-$) & $^{57}$Ti  & 13160\#&  0.068  & 34.93 ms &   22$\pm$2 ms\cite{half} &              &100\\
$^{58}$Sc($3^+$)   & $^{58}$Ti &16240\# &  0.269   & 11.03 ms  &12$\pm$5 ms\cite{half}    &              &100\\
$^{59}$Sc($7/2^-$) & $^{59}$Ti & 15340\#& 0.380     & 5.10 ms &10 ms($>$620 ns)\cite{half}&             &?\\
$^{60}$Sc($3^+$)   & $^{60}$Ti & 18280\#& 0.935     & 2.35 ms&3 ms($>$620 ns)\cite{half} &              &?\\
$^{61}$Sc($7/2^-$) & $^{61}$Ti & 17280\#&  0.388    & 3.31 ms&2 ms($>$620 ns)\cite{half} &              & ?\\
$^{56}$Ti($0^+$)   & $^{56}$V  & 6920&   0.661   & 224.5 ms&   200$\pm$5 ms\cite{half} &               &100\\
$^{57}$Ti($5/2^-$) & $^{57}$V  & 10360&  0.306    & 91.1 ms  &   95$\pm$6 ms\cite{half}  &             &100\\
$^{58}$Ti($0^+$)   & $^{58}$V  & 9210\#&    0.329  & 15.50 ms &   57$\pm$10 ms\cite{ENSDF}  &          &100\\
$^{59}$Ti($5/2^-$) & $^{59}$V & 12190\#&    0.1025  & 30.66 ms&   27.5$\pm$25 ms\cite{ENSDF}  &      &100\\
$^{60}$Ti($0^+$)   & $^{60}$V  & 10910\#&  0.586    & 3.697 ms & 22.2$\pm$25 ms\cite{ENSDF}&           &100 \\
$^{61}$Ti($1/2^-$) & $^{61}$V  & 14160\#&  0.207    & 9.07 ms  &15$\pm$4 ms\cite{half}    &           & 100\\
$^{62}$Ti($0^+$)   & $^{62}$V  & 12910\#& 1.138     & 1.029 ms &10 ms($>$620 ns)\cite{half}&               &100\\
$^{56}$V($1^+$)    & $^{56}$Cr & 9160&  0.230    & 206.32 ms &  216$\pm$4 ms\cite{half} &              & 100 \\
$^{57}$V($3/2^-$)  & $^{57}$Cr & 8300 &  0.476    & 513.7 ms  & 320$\pm$3 ms\cite{ENSDF}&           &100\\
$^{58}$V($1^+$)    & $^{58}$Cr & 9210\#&  0.230    &36.08 ms  &  191$\pm$10 ms\cite{half}&               &100\\
$^{59}$V($5/2^-$)  & $^{59}$Cr & 10060&  0.512    & 91.092 ms & 97$\pm$2 ms\cite{ENSDF}   &               &100\\
$^{60}$V($3^+$)    & $^{60}$Cr & 13260&  0.049    &112.71 ms   &  122$\pm$18 ms\cite{half}&              &100\\
$^{61}$V($3/2^-$)  & $^{61}$Cr &14160\# &  0.512    &44.95 ms   &  52.6$\pm$42 ms\cite{ENSDF}&              &100\\
$^{62}$V($3^+$)    & $^{62}$Cr & 15420\#&  0.214    & 23.128 ms &33.5$\pm$20 ms\cite{half}&              &100 \\
$^{63}$V($7/2^-$)  & $^{63}$Cr & 13730\#&   0.343   & 10.89 ms  &19.2$\pm$24 ms\cite{ENSDF}&             & 100    \\
$^{59}$Cr($1/2^-$) & $^{59}$Mn & 7630 &  0.076    & 805.5 ms  & 1050$\pm$90 ms\cite{half}&               &100 \\
$^{60}$Cr($0^+$)   & $^{60}$Mn & 6460 &  0.623    &  225.23 ms& 490$\pm$10 ms\cite{half} &              &100 \\
$^{61}$Cr($5/2^-$) & $^{61}$Mn & 9290 &  0.490    & 166 ms    &243$\pm$11 ms\cite{half}   &              &100\\
$^{62}$Cr($0^+$)   & $^{62}$Mn & 7590\#& 0.678     & 14.25 ms  & 206$\pm$12 ms\cite{half} &              &100\\
$^{63}$Cr($1/2^-$) & $^{63}$Mn &11160 & 0.098     &106.5 ms   &129$\pm$2 ms\cite{half}   &              &100\\
$^{64}$Cr($0^+$)   & $^{64}$Mn &9530\# &  1.447    & 3.734 ms  &  42$\pm$2 ms\cite{ENSDF}  &             &100\\
$^{60}$Mn($1^+$)   & $^{60}$Fe & 8444 &  0.308    &  171.73 ms&  280$\pm$20 ms\cite{half}&             &100\\
$^{61}$Mn($5/2^-$) & $^{61}$Fe & 7178 &  0.291    &  325.50 ms&  670$\pm$40 ms\cite{half}&             &100\\

\hline

\end{tabular}
%\label{be2}
\end{center}
\end{table}

%\onecolumn
\addtocounter{table}{-1}

\begin{table}
  \begin{center}
    \leavevmode
    \caption{{\em Continuation.\/}}   
\begin{tabular}{lrccccccccc}

% \begin{center} 
\hline
   &    & $Q$ value & Sum &\multicolumn{2}{c}{Half-life}&& $\beta^-$\\
\cline{5-6}
%\cline{8-9}   
$^{A}$Z$_{i}(J^{\pi})$  & $^{A}$Z$_{f}$  & (keV) &$B(GT)$&\multicolumn{1}{c}{Theo.}&\multicolumn{1}{c}{Expt.}&
   & $\%$\\
\cline{3-4}
%\cline{5-6}
%\cline{7-8}
\hline
$^{62}$Mn($1^+$)   & $^{62}$Fe & 10400\# &  0.391    &  952.6 ms&  92$\pm$13 ms\cite{half}&        &100\\
$^{63}$Mn($5/2^-$) & $^{63}$Fe & 8749 &  0.368    &  216.1 ms&  275$\pm$4 ms\cite{half}&          &100\\
$^{64}$Mn($1^+$)   & $^{64}$Fe & 11981&  0.271    &  60.53 ms&  90$\pm$4 ms\cite{Xu}&          &100\\
$^{65}$Mn($5/2^-$) & $^{65}$Fe & 10254&  0.436    &  73.21 ms&  84$\pm$8 ms\cite{Xu}&   &100\\
$^{65}$Fe($1/2^-$) & $^{65}$Co & 7964 &  0.880    &  69.59 ms&  810$\pm$50 ms\cite{half}&     &100\\
$^{66}$Fe($0^+$)   & $^{66}$Co & 6341 &  1.105    & 157.2 ms &  440$\pm$60 ms\cite{Xu}&             &100\\

$^{64}$Co($1^+$)   & $^{64}$Ni &7307  & 0.3481    & 241.8 ms & 300$\pm$30 ms\cite{half}    &         &100\\
$^{66}$Co($1^+$)   & $^{66}$Ni & 9598 &  0.300 & 132.35 ms &  200$\pm$2 ms\cite{Xu}   &           &100\\
$^{67}$Co($7/2^-$) & $^{67}$Ni & 8421 &   0.467& 99.98 ms&  425$\pm$20 ms\cite{Xu} &     &100\\

$^{67}$Ni($1/2^-$) & $^{67}$Cu & 3576 &  0.081 & 21.19 s &  21$\pm$1 s \cite{half}    & &  100        \\  
$^{68}$Ni($0^+$)   & $^{68}$Cu & 2103 &  1.079 & 61.32 s & 29$\pm$2 s\cite{half}   &    &         100      \\
$^{69}$Ni($9/2^+$) & $^{69}$Cu & 5758&  0.068 & 2.93 s &  11.5$\pm$0.3 s \cite{half}    & &  100        \\  
$^{70}$Ni($0^+$)   & $^{70}$Cu & 3763 &  0.670 & 3.42 s & 6.0$\pm$0.3 s\cite{half}   &    &         100      \\
$^{71}$Ni($9/2^+$) & $^{71}$Cu & 7305&  0.074 & 0.93 s &  2.56$\pm$0.03 s \cite{half}    & &  100        \\  
$^{72}$Ni($0^+$)   & $^{72}$Cu & 5557 &  0.674 & 0.51 s & 1.57$\pm$0.05 s\cite{half}   &    &         100      \\
$^{73}$Ni($9/2^+$) & $^{73}$Cu & 8879&  0.056 & 491.6 ms &  840$\pm$30 ms \cite{half}    & &  100        \\  
$^{74}$Ni($0^+$)   & $^{74}$Cu & 7550\#&  0.545 & 11.93 ms & 507.7$\pm$4.6 ms\cite{Xu}   &    &         100      \\
$^{75}$Ni($7/2^+$) & $^{75}$Cu &10230\#& 0.017 & 231.28 ms &  331.6$\pm$3.2 ms\cite{Xu}   &   & 100        \\  
$^{76}$Ni($0^+$)   & $^{76}$Cu & 9370\#& 0.282 & 9.904 ms & 234.6$\pm$2.7 ms\cite{Xu}      &  &   100      \\
$^{77}$Ni($9/2^+$) & $^{77}$Cu &11770\#& 0.018  &121.578 ms &  158.9$\pm$4.2 ms\cite{Xu}   &   & 100        \\  
$^{78}$Ni($0^+$)   & $^{78}$Cu & 10370\#& 0.870 & 2.255 ms & 122.2$\pm$5.1 ms\cite{half}  &   &   100      \\

$^{68}$Cu($1^+$)   & $^{68}$Zn &4440 & 0.077 & 36.12 s &  30.9$\pm$0.6 s\cite{half}     &  & 100        \\  
$^{69}$Cu($3/2^-$) & $^{69}$Zn & 2681& 0.043 & 2.0 m & 2.85$\pm$0.15 m\cite{half}     &   &   100      \\
$^{70}$Cu($6^-$)   & $^{70}$Zn &6588 & 0.035 & 3.05 s &  44.5$\pm$0.2 s\cite{half}     &  & 100        \\  
$^{71}$Cu($3/2^-$) & $^{71}$Zn &4618 & 0.029 & 19.9 s & 19.4$\pm$1.4 s\cite{half}     &   &   100      \\
$^{72}$Cu($2^-$)   & $^{72}$Zn &8363 & 0.037 & 9.8 s &  6.63$\pm$0.03 s\cite{half}     &  & 100        \\  
$^{73}$Cu($3/2^-$) & $^{73}$Zn &6606 & 0.035 & 3.58 s & 4.2$\pm$0.3 s\cite{half}     &   &   100      \\
$^{74}$Cu($2^-$)   & $^{74}$Zn &9751 & 0.013 & 1.36 s &  1.63$\pm$0.05 s\cite{half}     &  & 100        \\  
$^{75}$Cu($5/2^-$) & $^{75}$Zn &8088 & 0.057 & 1.11 s & 1.22$\pm$0.003 s\cite{half}     &   &   100      \\
$^{76}$Cu($3^-$)   & $^{76}$Zn &11327& 0.031 & 344.75 ms &  637.7$\pm$5.5 ms\cite{half}     &  & 100        \\  
$^{77}$Cu($5/2^-$) & $^{77}$Zn & 10280\#& 0.058 & 116.47 ms & 467.9$\pm$2.1 ms\cite{half}     &   &   100      \\
$^{78}$Cu($6^-$)   & $^{78}$Zn &12990& 0.011 & 541.4 ms &  330.7$\pm$2.0 ms\cite{Xu}    &       & 100        \\  
$^{79}$Cu($5/2^-$) & $^{79}$Zn & 11530\#& 0.019 & 164.3 ms & 241.3$\pm$2.1 ms\cite{Xu}    &   &       100      \\

$^{73}$Zn($1/2^-$) & $^{73}$Ga &4106& 0.296  & 36.14 s &  23.5$\pm$1 s\cite{half}    &  & 100        \\  
$^{74}$Zn($0^+$)   & $^{74}$Ga & 2293&0.416 & 27.9 s & 95.6$\pm$1.2 s\cite{half}       &&        100      \\
$^{75}$Zn($7/2^+$) & $^{75}$Ga &5906& 0.019  & 11.70 s &  10.2$\pm$0.2 s\cite{half}    &  & 100        \\  
$^{76}$Zn($0^+$)   & $^{76}$Ga & 3994&0.264 & 3.93 s & 5.7$\pm$0.3 s\cite{half}       &&        100      \\
$^{77}$Zn($7/2^+$) & $^{77}$Ga &7203& 0.068  & 1.10 s &  2.08$\pm$0.05 s\cite{half}    &  & 100        \\  
$^{78}$Zn($0^+$)   & $^{78}$Ga & 6223&0.267 & 0.51 s & 1.47$\pm$0.15 s\cite{half}       &&        100      \\
$^{79}$Zn($9/2^+$) & $^{79}$Ga &9115.4& 0.056  & 450.88 ms &  995$\pm$19 ms\cite{half}    &  & 100        \\  
$^{80}$Zn($0^+$)   & $^{80}$Ga & 7575&0.417 & 133.1 ms & 562.2$\pm$3.0 ms\cite{Xu}       &&        100      \\

\hline

\hline
\end{tabular}
\end{center}

\end{table}

\begin{figure}
\begin{center}
\includegraphics[width=14.2cm]{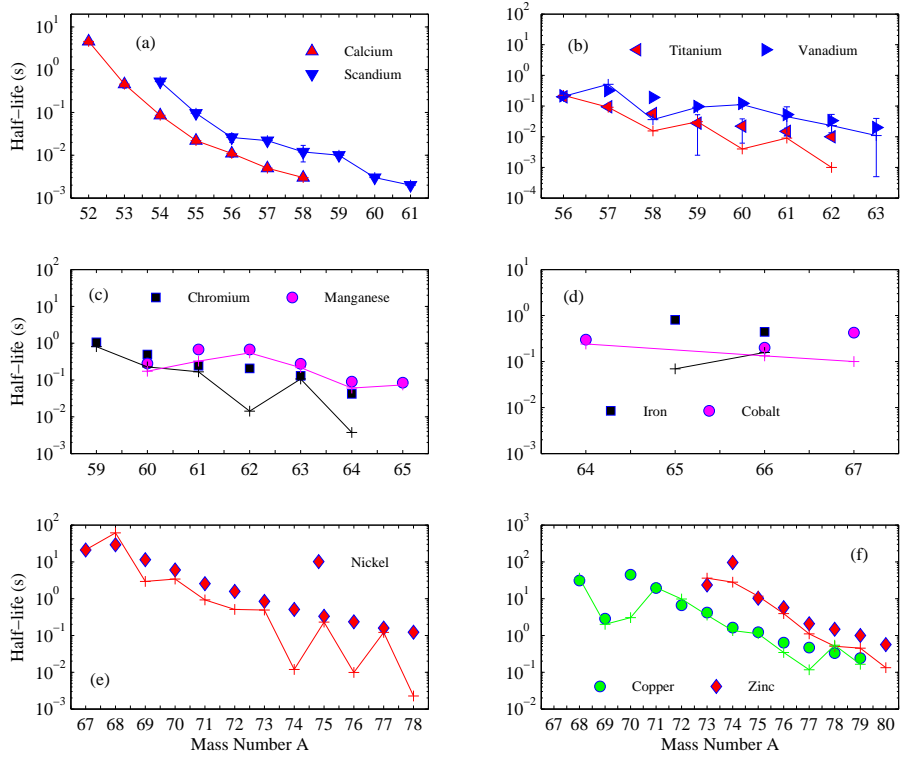}
\caption{ The $\beta$-decay half-life versus the mass number A of the concerned nuclei. The experimental results are
represented by points with error bars while the theoretical results are represented by solid lines with `+' sign. The nuclei are grouped
 by proton number $Z$.}
\end{center}
\end{figure}

\begin{figure}
\begin{center}
\includegraphics[width=16.2cm]{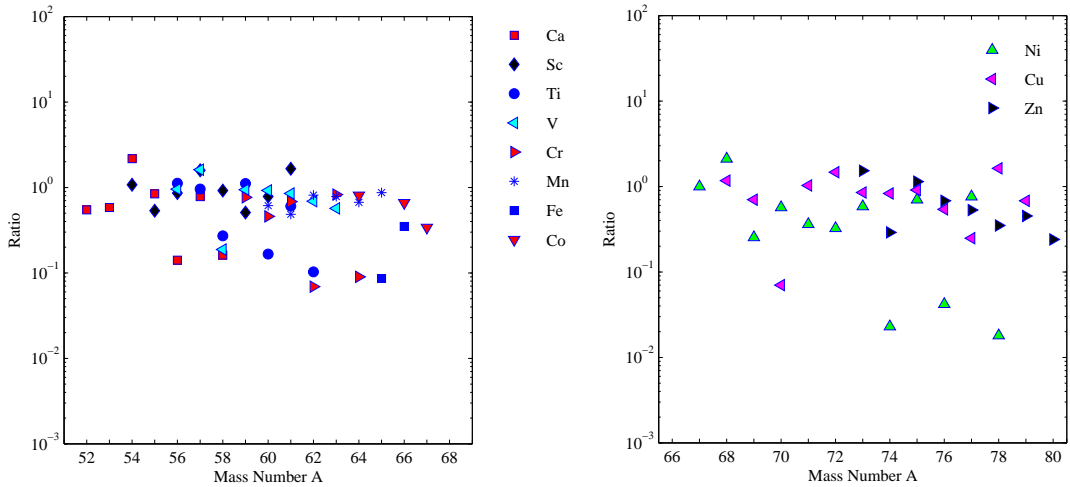}
\caption{ The ratios of theoretical to experimental half-lives versus the mass number $A$ for the concerned nuclei.}
\end{center}
\end{figure}

\begin{table}
  \begin{center}
    \leavevmode
    \caption{Comparison of the theoretical $\beta$-decay $Q$ values with the experimental data. The experimental data are taken from \cite{wong} where $`{\#}'$
    indicates that the presented value is estimated from systematic trends.}
    \label{tab:mgt_comp}
    \small
    \begin{tabular}{c@{\hspace{2pt}}c@{\hspace{2pt}}c@{\hspace{2pt}}c@{\hspace{2pt}}c@{\hspace{2pt}} c@{\hspace{2pt}}c@{\hspace{2pt}}c@{\hspace{2pt}}c@{\hspace{2pt}}c@{\hspace{2pt}}c@{\hspace{2pt}}c}
    \hline
     \multicolumn{3}{c}{$Q$ value (keV)}& & \multicolumn{3}{c}{$Q$ value (keV)} & & \multicolumn{3}{c}{$Q$ value (keV)} \\
     \cline{1-3}
     \cline{5-7}
     \cline{9-11}
     \multicolumn{1}{c}{Nuclei} & \multicolumn{1}{c}{Theo.}& \multicolumn{1}{c}{Expt.}&&
     \multicolumn{1}{c}{Nuclei} & \multicolumn{1}{c}{Theo.}& \multicolumn{1}{c}{Expt.}&&
     \multicolumn{1}{c}{Nuclei} & \multicolumn{1}{c}{Theo.}& \multicolumn{1}{c}{Expt.}\\
      \hline
      $^{52}$Ca &5795 & 5900$\pm$140&& $^{53}$Ca &9055 & 9650$\pm$480\#  &&$^{54}$Ca &9432 & 8820$\pm$620\# \\
      $^{55}$Ca &11867 & 11630$\pm$680\#&& $^{56}$Ca &11829 & 10830$\pm$720\#  &&$^{57}$Ca &14308 & 13830$\pm$780\# \\
      $^{58}$Ca &13934 & 12960$\pm$920\#&& $^{54}$Sc &10634 & 12000$\pm$380 &&$^{55}$Sc &10545 & 11690$\pm$490 \\
      $^{56}$Sc &13191 & 14470$\pm$420\# && $^{57}$Sc &12969 & 13160$\pm$560\# && $^{58}$Sc &15194 &16240$\pm$720\# \\
      $^{59}$Sc &14644 &15340$\pm$720\#&& $^{60}$Sc &17203 &18280$\pm$860\# &&  $^{61}$Sc &16426 & 17280$\pm$1000\#\\
      $^{56}$Ti &6752 & 6920$\pm$200 && $^{57}$Ti &9615 & 10360$\pm$330 && $^{58}$Ti &9413 & 9210$\pm$420\# \\
      $^{59}$Ti &11996 & 12190$\pm$430\#&&$^{60}$Ti &11672 & 10910$\pm$550\# && $^{61}$Ti &14151 & 14160$\pm$1080\# \\
      $^{62}$Ti &13631 & 12910$\pm$760\#&&$^{56}$V &7856 & 9160$\pm$180 &&$^{57}$V &7320 & 8300$\pm$230\\
      $^{58}$V &10667 & 9210$\pm$240&& $^{59}$V &10206 & 10060$\pm$290&&$^{60}$V &12581 & 13260$\pm$310\\
      $^{61}$V &12311 & 14160$\pm$900\#&& $^{62}$V &14801 & 15420$\pm$330\#&&$^{63}$V &14182 & 13730$\pm$610\#\\
      $^{59}$Cr & 5903  &7630$\pm$240&&$^{60}$Cr  & 5569  &6460$\pm$210 &&$^{61}$Cr  & 9480  &9290$\pm$130 \\
      $^{62}$Cr  & 9184  &7590$\pm$210\#&& $^{63}$Cr  & 11779  &11160$\pm$460&&$^{64}$Cr  &  11381 &9530$\pm$300\# \\
      $^{60}$Mn  &  6632 &8444$\pm$4 &&$^{61}$Mn  & 5646  &7178$\pm$3  &&$^{62}$Mn  & 8768  &10400$\pm$150\#\\
      $^{63}$Mn  &  9934 &8749$\pm$6   &&$^{64}$Mn  &  12181 &11981$\pm$6  &&$^{65}$Mn  &  11650 &10254$\pm$8\\
      $^{65}$Fe  & 8789&7964$\pm$7 && $^{73}$Ni  & 7799  &8879$\pm$3 &&$^{74}$Ni  & 5670  &7550$\pm$400 \\
      $^{75}$Ni & 9083  &10230$\pm$300\# &&$^{76}$Ni  & 7170  &9370$\pm$500\# &&$^{77}$Ni  & 10460  &11770$\pm$530\# \\
      $^{78}$Ni  & 8465  &10370$\pm$950\#  &&$^{76}$Cu  & 10380  &11327$\pm$7  &&$^{77}$Cu  & 8223  &10280$\pm$150\# \\
      $^{78}$Cu  & 11531  &12990$\pm$500\#&& $^{79}$Cu  & 9365  &11530$\pm$400\#&&$^{79}$Zn  & 8132  &9115.4$\pm$2.9\\
      $^{80}$Zn  & 6192  &7575$\pm$4 && &&\\

\hline
\end{tabular}
\end{center}
\end{table}

\begin{figure}
\begin{center}
\includegraphics[width=11.2cm]{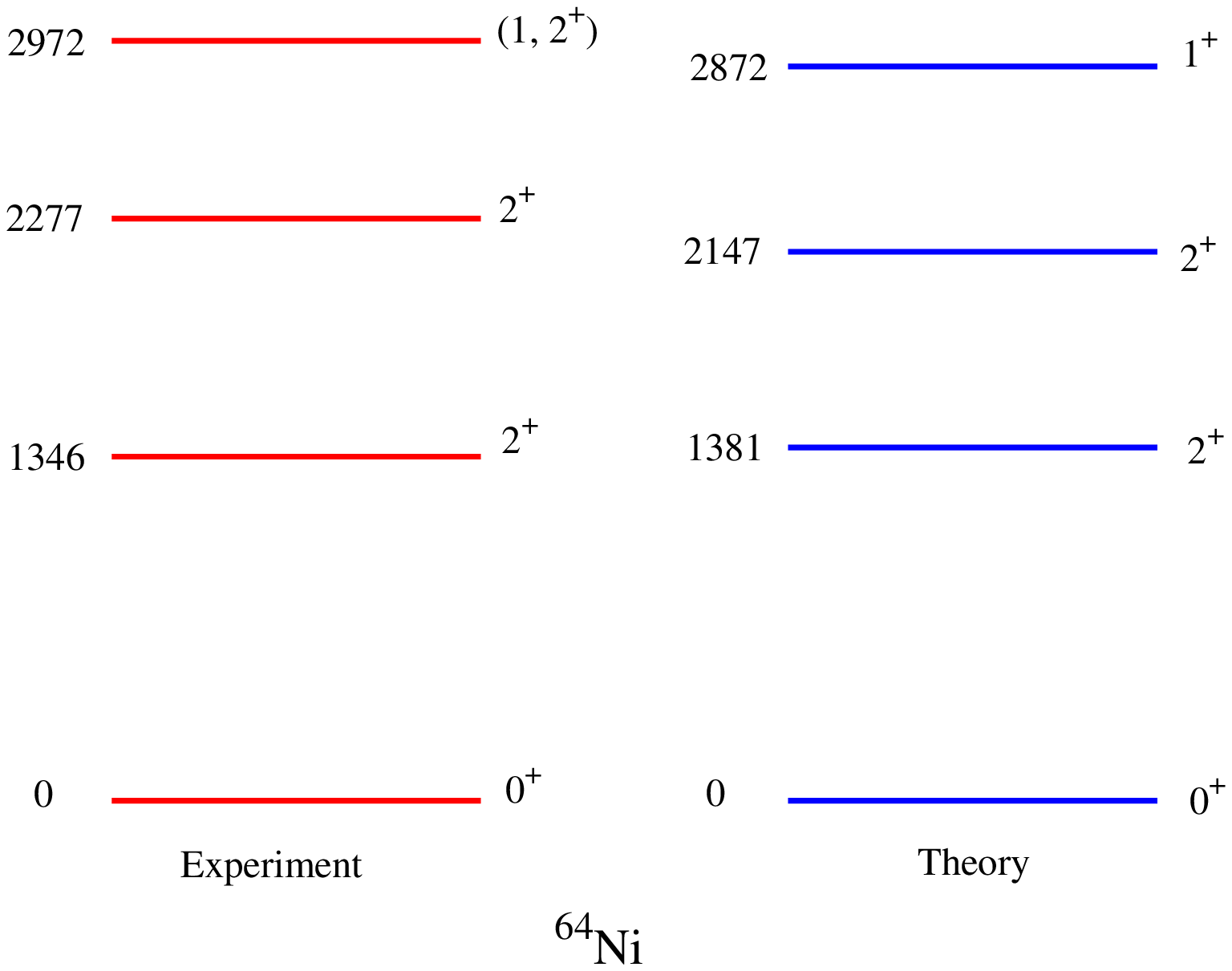}
\caption{ Comparison of the theoretical and experimental energy levels of $^{64}$Ni.}
\end{center}
\end{figure}

\section{Summary}

In the present work, we reported the half-lives, log$ft$ values, and branching fractions
of nuclei with $Z = 20 -30$ and N $\leq$ 50 using the nuclear shell model. We performed the calculation for $fp$ shell nuclei
using KB3G effective interaction.
In the case of Ni, Cu and Zn for $f_{5/2}pg_{9/2}$ model space we used JUN45 effective interaction.
The comparison with experimental results of
excitation energies, log$ft$  values, half-lives, $Q$-values and branching fractions for most of nuclei show
a good agreement with the available experimental data.
Some of the concerned nuclei belong to the island of inversion, such as $\mathrm{^{63-66}Mn}$ and $\mathrm{^{67}Co}$, and their $\beta$-decays are well reproduced by the calculations.
The present shell model results will add more information
to the earlier QRPA results \cite{q2}.
The QRPA results for half-lives are larger because it is pushing down GT-strength at low energies.
Further experimental half-lives measurement for very neutron-rich $fp$ and $f_{5/2}pg_{9/2}$ shell nuclei are strongly desired to
test shell model effective interaction.

%%%%%%%%%%%%%%%%%%%%%%%%%%%%%%%%%%%%%%%%%%%%%%%%%%%%%%%%%%

\section*{Acknowledgment:}

This work was supported in part by CSIR - India PhD fellowship. PCS acknowledges financial support from
faculty initiation grants.
H. Li acknowledges financial support from National Natural Science Foundation of China (grant numbers 11575082 and 11375086).

%%%%%%%%%%%%%%%%%%%%%%%%%%%%%%%%%%%%%%%%%%%%%%%%%%%%%%%%%%

\section*{References}
%\bibliographystyle{jphysg}
%\bibliography{refn82}

\end{document}